\newif\ifabstract
\abstracttrue   % for abstract
\abstractfalse   % for full version

\ifabstract  % *********************************

\documentstyle[twocolumn]{article}
\makeatletter
\def\section{\@startsection {section}{1}{\z@}{-3.5ex plus -1ex minus
      -.2ex}{2.3ex plus .2ex}{\large\bf}}
\makeatother

\else  % **********************************

\documentstyle[12pt]{article}

\fi % **********************************

\def\8{\infty}

\def\i{\imath\,}

\def\undertext#1{\vtop{\hbox{#1}\kern 1pt \hrule}}

\def\VEV#1{\left\langle\,#1\,\right\rangle}

\def\br{\\ \nonumber & &}

\def\be{\begin{equation}}
\def\ee{\end{equation}}
\def\bea{\begin{eqnarray} & &}
\def\eea{\end{eqnarray}}
\def\ct#1{\cite{#1}}
\def\rf#1{(\ref{#1})}

\def\t{\tilde}
\author{
{\sc Victor Gurarie}\\
  {\footnotesize Department of Physics}\\[-0.7ex]
  {\footnotesize Princeton University}\\[-0.7ex]
  {\footnotesize Princeton, NJ 08544}
} 
\title{Exact Computations \\ in the Burgers Problem}
\date {September 9, 1996}

\begin{document}
 \maketitle
\begin{abstract}

We complete the program outlined in the paper of the author  with
A. Migdal and  sum up exactly all the fluctuations around the instanton
solution of the randomly large scale driven Burgers equation.
We choose the force correlation function $\kappa$ to be exactly
quadratic function of the coordinate difference.
The resulting probability distribution satisfy the differential
equation proposed by Polyakov without an anomaly term.
The result shows that unless the anomaly term is indeed absent
it must come from other possible instanton solutions, and not
from the fluctuations.
\end{abstract}

In paper \ct{GM} we  computed the probability distribution $P(\delta u,
r)$
of observing the velocity difference $\delta u$ at a distance $r$ 
for the randomly large scale driven Burgers equation
in the WKB approximation. The results obtained there coincided with the
one obtained earlier in \ct{Pol}. Naturally, we were
able to derive the probability distribution only for $\delta u \gg 0$,
the so-called ``right tail'' of $P$. We found
\be
\label{EQpropx}
P(\delta u,r) \approx \exp \left( - {u^3 \over r^3} \right) 
\ee
At the same time, the left tail, conjectured in \ct{Pol} to be
\be
\label{EQprop}
P(\delta u,r) \approx {r^{3 \over 2} \over {| \delta u |}^{5 \over 2}} , \ \ 
- \sqrt {\left\langle u^2 \right\rangle } \ll \delta u \ll 
 0
\ee
remained to be computed. 

To compute the left tail, we have to go beyond the leading instanton
approximation. In this letter we show how to sum up all the
fluctuations around the instanton found in \ct{GM} in an exact way.
We find that the effect of the fluctuations around the instanton
is equivalent to a simple quantum mechanics (found previously
in \ct{Pol}) and therefore the summation is reduced to
much simpler problem of
solving that quantum mechanics. 

As was discussed in \ct{GM} the randomly driven Burgers equation is 
equivalent to the field theory with the action
\be
\label{EQactionold}
 S= - \i \int \mu (
u_t+u u_x -\nu u_{xx})+ {1 \over 2} \int dx dy dt  \ \mu(x,t)
\kappa(x-y) \mu(
y,t)
\ee
with coordinates $u(x)$
and momenta $\mu(x)$, the probability distribution of observing
the velocity profile $u(x)$ being the wave function in the
coordinate representation for \rf{EQactionold}. 

In \ct{GM} the wave function in the momentum representation was
constructed in the WKB, or instanton, approximation. 
Long range force-force correlation  was considered, 
\be
\VEV{f(x,t) f(y,t')}  = \delta(t-t') \kappa(x-y)
\ee
$\kappa$ is assumed to be a slow varying even function of $x$
which behaves as
\be
\label{EQforkappa}
\kappa(x)\approx \kappa(0)-{\kappa_0 \over 2} x^2, \ \
|x| \ll
\sqrt { \kappa(0) \over \kappa_0 } \equiv L
\ee
and quickly turns into zero when $|x| \gg L$.

While
only special wave functions in the momentum representation
were considered, let us for completeness give
a more general instanton solution. It is rather easy to
generalize the instanton found in \ct{GM} 
slightly to see that any correlation function
$\left\langle \exp \left( \int dx \lambda(x) u(x) \right) \right\rangle$
can be found as long as $\int dx \lambda(x) =0$. 
The velocity retains its linear profile and the answer is 
just\footnote { The construction of \ct{GM} is equivalent to taking
$\lambda(x) = \lambda_0 \left( \delta ( x - x_0/2) - \delta (x +
x_0/2) \right)$}
\be
\label{Ebigans}
\left\langle \exp \left( \int dx \lambda(x) u(x) \right) \right\rangle =
\exp \left\{ { \sqrt {2 \kappa_0} \over 3}
{ \left[  \int dx x \lambda(x) \right] }^{3 \over 2} 
\right\}, \ \ \int dx x \lambda(x) \gg 0
\ee 
The support of $\lambda(x)$ must lie within the $\pm L$ interval.

The  natural thing to do now would be to consider the fluctuations
around the instanton solution, thus computing the corrections to the
answer \rf{Ebigans}. That was the program outlined in \ct{GM}.

In principle, the next order correction to \rf{Ebigans} is given by
the determinant of the operator we derive by expanding the action
\rf{EQactionold} around the instanton solution. 
In practice, 
to compute the determinant directly would be rather inconvenient. 
So we take a slightly different approach suggested in \ct{GM}. 
Instead of computing the quantity\footnote{We introduce an auxiliary
parameter $\lambda_0$ which we can later set to 1.}
\be
\label{EQtony}
Z[\lambda_0 \lambda(x)] =\left\langle \exp 
\left( \int dx \lambda_0 \lambda(x) u(x) \right) \right\rangle
\ee
we introduce
\be
\label{EQpert}
F={d \over d \lambda_0} \log Z[\lambda_0 \lambda(x)] 
\ee
This quantity, as was shown in \ct{GM}, is easy to expand around the
instanton solution. We choose
\bea
u = u_{\rm inst} + \t{u} \br
\mu = \mu_{\rm inst} + \t{\mu} 
\eea
$u_{\rm inst}$ and $\mu_{\rm inst}$ being the instanton solution, and find
\be
\label{EQwhat}
F=\int dx \lambda(x) u_{\rm inst}(x) + { \int dx \lambda(x)
 \int {\cal D} \t{u} {\cal D} \t{\mu} \ \t{u}(x) \exp \left( -\t{S} \right)
\over \int {\cal D} \t{u} {\cal D} \t{\mu} \ \exp \left( -\t{S} \right)
}
\ee

The action $\t{S}$ is the expansion 
of the  action \rf{EQactionold},
\bea
\label{EQact}
\t{S} = - \i \int \t{\mu} (
\t{u}_t -\nu \t{u}_{xx})+ {1 \over 2} \int dx dy dt  \ \t{\mu}(x,t)
\kappa(x-y) \t{\mu}(y,t) + \br + \i \int  
\t{\mu}_x {\partial \over \partial x} \left( \t{u}
 u_{\rm inst}
\right) 
- \i \int \mu_{\rm inst} \t{u} \t{u_x} 
- \i \int \t{\mu} \t{u} \t{u_x}
\eea
The last term of \rf{EQact} is the interaction. We will refer
to the action
{\sl without} the interaction term as $\t{S}_0$.

The next step is to define the Green's functions of small fluctuations around
the instanton. These are the Green's function of the free field theory
provided for us by $\t{S}_0$. For example, if we define
\bea
\label {EQGreens}
G(x,t; x',t') = \i {\left\langle \t{\mu}(x,t) \t{u}(x',t') \right\rangle}_0
\br
N(x,t; x',t') = {\left\langle \t{u}(x,t) \t{u}(x',t') \right\rangle}_0
\br
M(x,t; x',t') = - {\left\langle \t{\mu}(x,t) \t{\mu}(x',t') \right\rangle}_0
\eea
the brackets $\langle \  \rangle_0$ signifying the fact that we define these
functions with the action $S_0$.
We can easily find the equations these Green's functions satisfy,
for instance
\bea
\label{EQGreq}
N_t +{\partial \over \partial x} \left( u_{\rm inst} N \right) 
- \nu N_{xx} = - \int dy \ \kappa(x-y) \ G(y,t; x',t')
\br
G_t + u_{\rm inst} {\partial G \over \partial x} + \nu G_{xx} =
- \i {\partial \mu_{\rm inst} \over \partial x} N +
\delta (x-x') \delta (t-t')
\eea
We could try to solve these equations, find the Green's functions
and then start expanding \rf{EQwhat}  in powers of the interaction.
The perturbation theory (complete with Feynman diagrams) we get in this
way is actually a perturbation theory in powers of $1/\lambda$. 
In that sense, it is much better defined than the perturbation theory
we would get if we expanded the action around the zero value of the fields,
as in the standard Wyld's technique (see \ct{Wyl}).

Using the perturbation theory around the instanton we could construct
$F$ as a function of $\lambda_0$ and then we could integrate it
to reconstruct the quantity $Z$.

However, to solve the equations \rf{EQGreq} in the
straightforward way is difficult. If, for instance, we used the expansion 
\rf{EQforkappa} we 
could try to expand the function $N$ in powers
of $x$ and $x'$ as well,
\be
\label{EQunstable}
N(x,t;x',t') =  a(t,t') + b(t,t') x x' + c(t,t') (x^2 + {x'}^2)
\ee
and try to find the functions $a$, $b$ and $c$. This approach would
not work as $a$ and $c$ will diverge at $t \rightarrow
- \infty$. That shows that the function $N$ is more complicated than
just a simple polynomial.  

That is why we are going to use a certain trick which will allow us
to avoid solving \rf{EQGreq}.

We begin with writing down the instanton equations of motion again,
\be\label{EQmotiono}
 u_t + u u_x - \nu u_{xx}  = - \i \int dy  \
\kappa (x-y)
\mu(y)
\ee
\be
\label{EQmotiont}
\mu_t + u \mu_x + \nu \mu_{xx}  = 
0
\ee

These equations, since they follow from the action \rf{EQactionold}, 
have at least
three conservation laws. The first two are the generalized momentum and
energy, the corresponding components of the energy-momentum tensor,
\bea
E= \i \int dx \mu (u u_x - \nu u_{xx}) - {1 \over 2} \int dx dy \ \mu(x)
\kappa(x-y) \mu(y)
\br
P = \int dx \mu u_x
\eea
They were originally found in \ct{LF}. In addition to these two,
there is another conservation law  following from the Galilean
invariance
of \rf{EQactionold}. The transformation laws are
\bea
\label{EQGalileo}
u(x) \rightarrow u(x- v t) +v \br
\mu(x) \rightarrow \mu(x- vt)
\eea
and the corresponding conserved quantity
\be
G=\int dx \mu - t P
\ee
If the generalized momenum $P$ is zero (and it will be zero for all the
solutions of \rf{EQmotiono} and \rf{EQmotiont} which 
fall to zero at $t \rightarrow -\infty$), then even the quantity
\be
\t{G}=\int dx \mu
\ee
is conserved and zero. The quantity $\t{G}$ can be shown to be a 
canonical momentum conjugate to the constant component of the
velocity field. It has an evident meaning of the center of mass
of Burgers fluid (moving with the velocity given by the momentum $P$).

It is important to note  that 
the Galilean symmetry is, in fact, spontaneously broken since
\be
\VEV{u^2} = {\rm const}
\ee
as was pointed out in \ct{Pol}.
Spontaneous symmetry breaking is, as always, due to the
boundary conditions in path integral. The natural boundary
condition is $u(t=-\infty)=0$. If we take $u(t=-\infty)=v$,
we obtain another ground state of the system which can be 
taken to the first one by the symmetry transformation  \rf{EQGalileo}. 
By the way, the instanton found in \ct {GM} satisfied the boundary
condition  $u(t=-\infty)=0$ and therefore it was not invariant under
\rf {EQGalileo}; by applying \rf{EQGalileo} to it we could construct
other instantons. We did not need to sum over all those instantons,
though, because the symmetry is broken.
Fortunately, this spontaneous symmetry breaking does not prevent
us from exploring the consequences of the Galilean symmetry.

We saw that the Galilean symmetry constraints 
the momentum conjugate
to the constant component
of the velocity  which was the
culprit of the unstability in the Green's functions (compare with
\rf{EQunstable}).
Could we find a way to constraint the 
integral over the momentum $\int dx  \mu$ not only on classical
solutions, but also in all the fluctuations? It turns out there
is such a way.

Following \ct {LF}, we consider a slight modification of
\rf{EQGalileo} allowing $v$ to depend on time. This is no longer a
symmetry of the action \rf {EQactionold}. Under the 
proposed transformation
\bea
\label{EQFFF}
u(x) \rightarrow u\left( x-r(t) \right) + {d r(t) \over dt }
\br
\mu(x) \rightarrow \mu \left( x - r(t) \right) \br
r(0)=0, \ \ {d r \over d t}(t=-\infty) =0
\eea
the action \rf {EQactionold} changes as
\be
S \rightarrow S - \i \int dt \left\{ 
{d^2 r \over dt^2} \int dx \mu  \right\}
\ee
Note that the transformation does not change the ground state
due to the last condition in \rf{EQFFF}.

It turns out that there is a procedure first proposed in \ct{LF} 
(for a diferent reason) which is similar to the 
computation of the Faddeev-Popov  determinant and which
allows to integrate out the
degree of freedom associated with $r(t)$. We would like to
give here the outline of this procedure.  We begin with inserting into the
functional integral $\int {\cal D} u {\cal D} \mu \exp(-S)$
the identity
\be
\int {\cal D} r \delta \left( u(r) - {d r \over d t} \right) 
J[u] = 1
\ee
which serves as a definition of the quantity $J[u]$, the Faddeev-Popov
determinant. 
Then we shift the variables of integration according to \rf{EQFFF}
and integrate out $r(t)$
to obtain the original path integral 
with the Faddeev-Popov determinant and two constraints.
\be
\label{EQactionnew}
\int {\cal D}u {\cal D}\mu \exp(-S) \  J[u] \ \delta (u(0)) \ \delta 
\left( \int dx {d^2 \mu \over dt^2} \right)
\ee

To proceed further we need to compute the Jacobian $J[u]$. 
This Jacobian is just
\be
\label {EQJak}
J[u] = \det \left( {d \over dt} - {\delta u (r) \over \delta r} \right)
\ee
We have to understand that the operator whose Jacobian we are computing
defines the time propagation backwards; $r(0)=0$ and then we find
$r(t)$ for $t < 0$. That allows us to compute the Jacobian using the
standard methods (see, for example, \ct{Jus}) to get
\be
\label{EQZak1}
J[u] = \exp \left[ \theta(0) \int dt \ u_x (0)  \right]
\ee
In this form the formula is rather ambiguous
and depends on how we define the $\theta$-function at 0. 
The necessity to define the $\theta$ function at 0 is often encountered
in path integral formalism. 
It is related to the way we discretize time in deriving \rf{EQZak1}.
By discretizing time we break Galilean invariance and it is not
evident that it will be restored in the limit when we take the
discretization time to zero. We have to choose such discretization
so as to preserve Galilean invariance in the continuous limit.

Unfortunately, so far no one has managed to show by a direct computation
what the correct choice for $\theta(0)$ should be. However, 
in the paper \ct{Krai} an indirect procedure was developed which, 
when applied to the formalism in hand, supports the choice $\theta(0)=1$.
Such a choice would be equivalent to taking the retarded discretization
of Burgers equation and was already suggested in \ct{LF}. 
As we will see later, it is also consistent with the Polyakov's
equation. 

Therefore, the Jacobian must be taken as
\be
\label{EQJacc}
J[u] = \exp \left[ \int dt \ u_x (0)  \right]
\ee

Now note that
the two constraints precisely eliminate two unstable modes 
(compare with \rf{EQunstable}). 
The Jacobian \rf{EQJacc} is in fact an effective contribution
of the modes we eliminated in this way. 

Now we can safely go on and find the Green's functions. The equations for
the Green's function will be changed slightly to account for the constraints
we have. For example, the equations \rf{EQGreq} will become (we use
$u_{\rm inst} = \sigma x$)
\bea
\label{EQGreq1}
N_t + \sigma {\partial \over \partial x} \left( x N \right) 
- \nu N_{xx} = - \int dy \ \kappa(x-y) \ G(y,t; x',t') +
f(t; x',t')
\br
G_t + \sigma x {\partial G \over \partial x} + \nu G_{xx} =
- \i {\partial \mu_{\rm inst} \over \partial x} N +
[\delta (x-x') - \delta(x)]\delta (t-t')
\eea
the difference of the delta functions being necessary to ensure
$\int dx \ G=0$ and $G(x'=0)=0$ and the function $f(t; x',t')$
being chosen to ensure $N(x=0)=N(x'=0)=0$.

It turns out that unlike \rf{EQGreq} 
it is relatively easy to solve \rf{EQGreq1}. We expand the
function $\kappa$ in Taylor series
and immediately see that
the function $N$ is bilinear in $x$ and $x'$,
\be
N(x,t; x',t') = b(t,t') x x'
\ee
where the function $b(t,t')$ coincides {\sl exactly} with the
correlation of fluctuations of the velocity strain. The functions
$a$ and $c$ from \rf{EQunstable} disappear due to the constraints.
Moreover, 
by a direct computation one can show that the perturbative expansion
for the quantity $Z$ defined in \rf{EQtony} coincides term by term
with a perturbative expansion of the quantity
\be
Z=\left\langle \exp \left[ \sigma \lambda_0 \int dx x \lambda(x) 
\right] \right\rangle
\ee defined in a theory with the action
\be
S_{QM}=\int dt \left\{ 
i p (\sigma_t + \sigma^2) - {\kappa_0 \over 2} p^2 + 
\sigma \right\}
\ee
This action follows from \rf{EQactionnew} if we formally substitute
$u=\sigma(t) x$ and $p=\int dx \ x \mu(x)$. The last term $\sigma$ comes
out of the Jacobian \rf{EQJacc}.

$S_{QM}$ defines the motion in the potential\footnote{We have 
to  put
$p$ on the left from $\sigma$ in the
Schr\"odinger
equation, see \ct{Jus}.} 
\be
V(\sigma) = {\sigma^4 \over 2 \kappa_0} - 2 \sigma
\ee
This is the quantum mechanics 
discussed in \ct{Pol} without the anomaly term (in the notations
of \ct{Pol} $b=1$\footnote{Had we kept $\theta(0)$ in our formulas
we would have obtained ${\sigma^4 \over 2 \kappa_0} - (1+\theta(0)) \sigma$
showing that $b={1+\theta(0) \over 2}$. 
Therefore, a choice $\theta(0)={1 \over 2}$
is tempting. Unfortunately, it has not been derived in any way known
to the author so far.}).

This quantum mechanics was shown in \ct{Pol} not to have
a stationary state and gives an asymptotics to the (nonstationary)
probability distribution
\be
P(\delta u) \approx {1 \over  |\delta u|^{3}}, \ \ \delta u \ll 0
\ee
This is certainly not what is observed in the numerical experiments,
\ct{YC}. And we note that the viscosity did not contribute to the
calculations.

This result is physically unsatisfactory. We would have expected
to find the anomalous contribution to the Polyakov's equation
due to nonzero viscosity. One possible explanation could be that
the instanton found in \ct{GM} is not the only possible instanton
solution.
If this is so, then the further program is to find the instanton
solutions for Burgers problem which do not use the Taylor expansion
of the force correlation function. Perhaps they will contain
the shock waves combined with linear profiles \ct{Polumn} and
the fluctuations around them  will produce us the anomaly term
of \ct{Pol} and lead to the stationary state.

This problem notwithstanding, we would like to conclude with saying 
that we produced an answer for the Burgers problem in the
region where the shock waves are absent by summing up exactly 
the perturbations around the instanton found in \ct{GM}.

The author is grateful to A. Migdal who was a collaborator in this
project at its earlier stages, 
and to S. Boldyrev, A. Polyakov, E. Balkovsky,
V. Lebedev, G. Falkovich and I. Kolokolov
for
numerous and important discussions which  helped to
find the correct interpretation to the  results 
obtained in  this paper.

\end{document}